\documentstyle[prl,aps,epsf]{revtex}

\newcommand{\req}[1]{(\ref{#1})}
\newcommand{\beq}{\begin{equation}}
\newcommand{\eeq}{\end{equation}}
\newcommand{\beqar}{\begin{eqnarray}}
\newcommand{\eeqar}{\end{eqnarray}}
\newcommand{\pair}[2]
{$\left(\begin{array}{c} #1\\[-0.0cm]#2 \end{array}\right)$}

\begin{document}
\draft
\twocolumn[\hsize\textwidth\columnwidth\hsize\csname
@twocolumnfalse\endcsname
\title{$1/f$ Noise and Long Configuration Memory\\
in Bak-Tang-Wiesenfeld Models on Narrow Stripes}
\author{Sergei Maslov$^1$, Chao Tang$^2$, and Yi-Cheng Zhang$^3$}

\address{$^1$ Department of Physics, Brookhaven National Laboratory,
Upton, New York 11973}
\address{$^2$ NEC Research Institute, 4 Independence Way, Princeton, New
Jersey 08540}
\address{$^3$ Institut de Physique Th\'{e}orique, Universit\'{e} de
Fribourg, Fribourg CH-1700, Switzerland}

\date{February 5, 1999}
\maketitle
\begin{abstract}
We report our findings of an $1/f$ power spectrum for the total amount 
of sand in directed and undirected Bak-Tang-Wiesenfeld models confined
on narrow stripes and driven locally. The underlying mechanism for the
$1/f$ noise in these systems is an exponentially long configuration
memory giving rise to a very broad distribution of time scales. Both
models are solved analytically with the help of an operator algebra to
explicitly show the appearance of the long configuration memory.
\end{abstract}
]

\section{Introduction}
The ubiquitous $1/f$ noise fascinated physicists for generations
\cite{press,dutta}. There are many examples from wildly different
systems in which the power spectra $S(f) \sim 1/f^\alpha$
with $\alpha$ close to one.
This phenomenon usually indicates the presense of a broad
distribution of time scales in the system. The common case where
$\alpha=1$ is particularly intriguing, in that it implies a kind of
``equal partition'' of power among every decade of frequency range,
i.e. the integral
\begin{displaymath}
\int_{f}^{10f} S(f) df \sim \int_{f}^{10f} \frac{1}{f} df 
= \int_{f}^{10f} d \ln f = \ln 10 
\end{displaymath}
is independent of $f$.
One mechanism to generate such a distribution of time scales in
systems at thermal
equilibrium is through thermal activation events over a sufficiently
broad and flat distribution of energy barriers \cite{dutta}.
The ``local'' power spectrum generated by any single barrier has a
characteristic frequency which decreases exponentially with increasing
barrier height. But the superposition of the power spectra from all the
barriers gives rise to an $1/f$ spectrum.
This mechanism is often employed to explain, for example, the $1/f$
spectrum of low 
frequency voltage fluctuations in semi-conductors \cite{dutta}. 
In search for a more general answer, applicable to nonequilibrium and
dynamic systems, Bak, Tang, and Wiesenfeld (BTW)
introduced the notion of Self-Organized Criticality (SOC) \cite{btw}.
In particular, they proposed a simple ``sandpile'' (BTW) model which 
shows the emergent scale free behavior in both space and time.
However, the original BTW model did not exhibit the $1/f$ noise
\cite{jensen}. 
In this paper we report the observation of $1/f$ noise for directed
and standard (undirected) BTW models confined on narrow stripes 
(quasi-one-dimensional geometries). In these models, sand flows in the
long direction, with periodic or closed boundary conditions in the other
direction.
The system is driven locally by randomly adding sand to a unique set
of sites that have the same coordinate along the long axis.
The total amount of sand in the sandpile as a function of time,
measured by the number of added grains, exhibits a clean $1/f$ power
spectrum with an exponentially small lower  cutoff. 
Surprisingly, the mechanism for the $1/f$ noise in this 
athermal nonequilibrium model is rather similar to the above
mentioned thermal mechanism. In our model the local characteristic
frequency also falls off exponentially as a function of 
some parameter, which in this case is the distance from 
the driving point.

\section{Directed models}
Let us first consider the simpler directed model, defined as
follows.  An integer variable $z(x,y)$ is assigned on every site
$(x,y)$ of a two-dimensional lattice of size $L_x \times L_y$ ($1 \le x
\le L_x$, $1 \le y \le L_y$). Throughout the paper, we refer to $z(x,y)$
as the number of grains of sand (or height) at the site $(x,y)$.
The dynamics consists of the following steps:
\begin{itemize}
\item[(i)] Add a grain of sand to a randomly selected site in the {\it
first}
column, $(1,y)$: $z(1,y) \to z(1,y) +1$.
\item[(ii)] If as a result of the process the height $z(x,y)$ exceeds
a critical value $z_c=2$, the site topples and three grains of sand 
are redistributed from this site to three of its nearest 
neighbors up, down, and to the right, that is
\begin{eqnarray*}
z(x,y+1) &\to& z(x,y+1)+1,   \\
z(x,y-1) &\to& z(x,y-1)+1,   \\
z(x+1,y) &\to& z(x+1,y)+1,   \\
z(x,y) &\to& z(x,y) - 3. 
\end{eqnarray*}
\item[(iii)] Repeat step (ii) until all sites are stable, i.e. 
$z(x,y) \leq 2$ everywhere. This chain reaction of updates 
is referred to as an avalanche.
\item[(iv)] When the avalanche is over, measure the total amount of sand
in
the system ${\cal Z}(t)=\sum z(x,y)$. Then go to step (i).
\end{itemize}
Notice that the flow of sand is {\it directed} to the right along the
$x$-axis. Also note the separation of time scales -- the duration of
individual avalanches is taken to be much faster than the unit time
interval which is defined by the addtion of sand grains.
The boundary condition in the $x$ direction is always set to open:
$z(L_x+1,y)=0$. While in the $y$ direction we either choose the periodic
boundary condition (which we refer to as Model 1) or the closed boundary
condition. In the latter case we restrict ourselves to $L_y=2$ and refer
to it as Model 1A. In Model 1A, we set $z_c=1$ and the redistribution
rule (ii) prescribes to move two grains of sand from the toppling site: one
to the right along the $x$ direction and the other to its nearest
neighbor (up or down) in the $y$ direction.
\begin{figure}[h]
\narrowtext
\centerline{\epsfxsize=3.3in
\epsffile{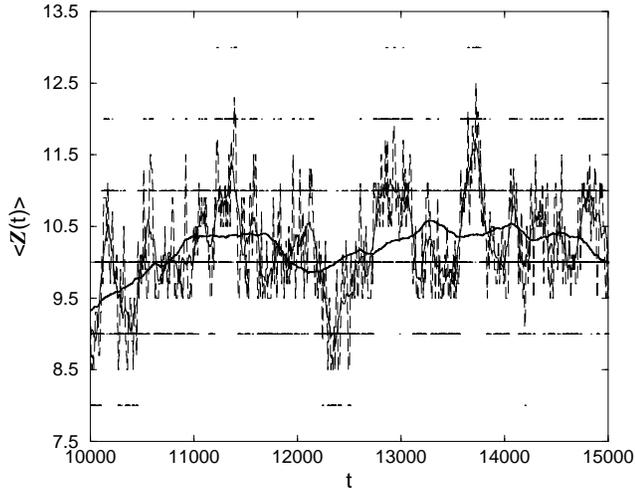}}
\caption{The total amount of sand in the system ${\cal Z}(t)$ (dots),
and its running averages over 10 (dashed line), 100 (solid line), and
1000 (thick solid line) time steps. The data was taken from Model 1A
with
$L_x=8$.}
\label{zt}
\end{figure}

\begin{figure}[h]
\narrowtext
\centerline{\epsfxsize=3.3in
\epsffile{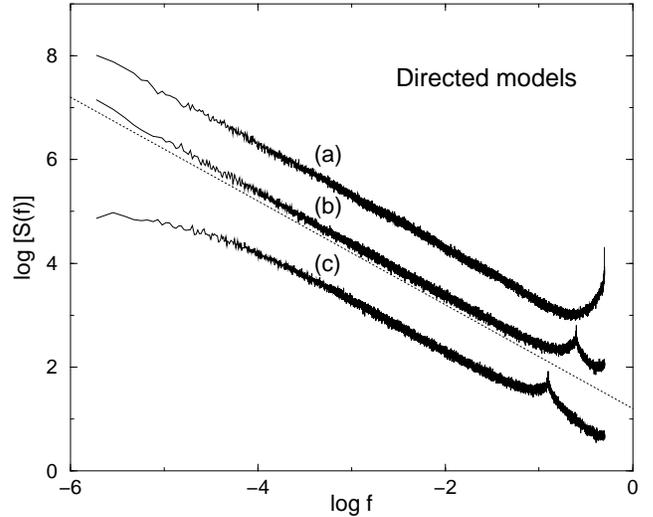}}
\caption{Power spectra for (a) Model 1A with $L_x=8$; (b)
Model 1 with $L_x=8$ and $L_y=4$; and (c) Model 1 with $L_x=8$ and
$L_y=8$. The curves in (b) and (c) are shifted vertically by -1 and -2
decades, respectively, for clarity. The dashed line has the slope $-1$.}
\label{pw-di}
\end{figure}

After some transient period the above dynamics brings the system to a
stationary state, where the total amount of sand in the system ${\cal
Z}$ saturates and fluctuates about its average value. At this point we
start recording ${\cal Z}(t)$ and measure its power spectrum 
$S(f)=|\hat{{\cal Z}}(f)|^2$, where $\hat{{\cal Z}}(f)$ is the
Fourier transform of ${\cal Z}(t)$. A time window of ${\cal Z}(t)$ is
shown in Fig.~\ref{zt} together with its running averages. Notice the
fluctuations on many time scales. In Fig.~\ref{pw-di},
we show the power spectra for Models 1 and 1A. 
Even for small systems one observes a very broad $1/f$ region. In fact,
as we will demonstrate later, the lower cutoff of the $1/f$ region 
falls off with $L_x$ exponentially. 
Our simulations indicate that as the width of the stripe in Model 1,
$L_y$ is increased the $1/f$ region shrinks, as shown in
Fig.~\ref{pw-di}(c).
Direct observations of configurational changes at each time step clearly
indicate that the rate of configurational changes at $x$ decreases
drastically with increasing $x$ \cite{java}, suggesting that there are
many time scales and some kind of long memory in the system. To
understand this, we proceed with solving Model 1A using the group of
operators introduced by Dhar \cite{dhar,dhar_review} to treat 
sandpile models.

To simplify the notation let us denote the configuration at the 
pair of sites $z(x,1)$ and
$z(x,2)$ by the column $\left( \begin{array}{c} z(x,2) \\[-0.0cm]
z(x,1) \end{array} \right)$, and let $L_x=L$.
Any pair configuration with both $z(x,y) \leq z_c = 1$ is stable.
However, the recurrent pair configurations, present in the
stationary SOC state, are $\left(\begin{array}{c} 0 \\[-0.0cm]
1\end{array} \right)$, $\left( \begin{array}{c}
1\\[-0.0cm]0\end{array}\right)$, and $\left(
\begin{array}{c} 1 \\[-0.0cm] 1 \end{array} \right)$, 
while $\left(\begin{array}{c} 0 \\[-0.0cm]
0 \end{array} \right)$ is never realized in the system after some
transient period. As usual in directed models \cite{dhar_ramaswamy}, 
there are no additional restrictions on pair configurations 
at different columns.
The total number of recurrent (SOC) states is thus $3^L$. Let us define
operators $U_x$ and $D_x$ acting on recurrent configurations. 
The action of these operators consists of adding one grain 
of sand at sites $(x,2)$ and $(x,1)$ correspondingly, and, if
necessary, relaxing the resulting configuration according to the
avalanche rules of the dynamics. The final stable configuration 
is the result of the operator acting on the initial configuration. 
One can demonstrate that these operators commute with each 
other, i.e. $[U_x,U_{x'}] = [D_x,D_{x'}] =[U_x,D_{x'}]=0$, 
and the model is therefore an Abelian model \cite{dhar}. 
The following operator identities \cite{dhar} result directly from 
the relaxation rules of the model 
\begin{eqnarray}
U_x^2 &=& D_x U_{x+1} \label{u2} \\
D_x^2 &=& U_x D_{x+1}.
\label{d2}
\end{eqnarray}
These identities simply state that the addition of two grains of sand
to any site (of a recurrent state) will certainly make it unstable 
and, therefore, two grains will be transferred to the neighbors
according to
the relaxation rule of Model 1A.
Open boundary at $x=L+1$ corresponds to $U_{L+1}=D_{L+1}={\bf I}$,
where ${\bf I}$ is the identity operator. From 
Eqs. (\ref{u2}) and (\ref{d2}) it immediately follows that
\begin{equation}
U_1D_1=U_2D_2= \cdots =U_LD_L=U_{L+1}D_{L+1}={\bf I}.
\label{ud}
\end{equation}
In other words, addition of one grain of sand to both upper and lower
sites at any column $x$ triggers a downstream avalanche, in which two
grains fall off the right edge, but the underlying configuration of
sand columns remains unchanged. Since $D_x=U_x^{-1}$, 
Eq.~(\ref{u2}) can be rewritten as $U_x^3=U_{x+1}$. 
Therefore, $U_1^{3^L}=U_{2}^
{3^{L-1}}=U_{L}^{3}=U_{L+1}$, and
\begin{equation}
U_1^{3^L}={\bf I}.
\label{u3L}
\end{equation}
Repeated application of $U_1$ $3^L$ times makes the system visit every
one of the $3^L$ recurrent states exactly once and return to its
original configuration. If $U_1$ and $D_1$ are applied in random
order (that is how we drive the system in Model 1A), 
eventually all recurrent states will still be visited. But, since
$U_1^mD_1^n=U_1^{m-n}$, for a random sequence of $U_1$'s and $D_1$'s 
the average time required to visit all $3^L$ states
is given by $\sqrt{T}=3^{L}$, or $T=9^L$.
Now let us study the ``microscopic'' details of how operators $U_x$ and
$D_x$ change the configurations of the sandpile. Note that
\beqar
U_{x}\left(\begin{array}{c} 1\\[-0.0cm]0 \end{array}\right)&=&\left(
\begin{array}{c} 2\\[-0.0cm]0 \end{array}\right)=\left(\begin{array}{c}
0\\[-0.0cm]1 \end{array}\right)U_{x+1}, \label{u10} \\
U_{x}\left(\begin{array}{c} 0\\[-0.0cm]1 \end{array}\right)&=&\left(
\begin{array}{c} 1\\[-0.0cm]1 \end{array}\right), \label{u01} \\
U_{x}\left(\begin{array}{c} 1\\[-0.0cm]1 \end{array}\right)&=&
\left(\begin{array}{c}
0\\[-0.0cm]2 \end{array}\right)U_{x+1}=\left(\begin{array}{c}
1\\[-0.0cm]0 \end{array}\right){\bf I}. 
\eeqar
It is clear that the only cases where the operator $U_x$ or $D_x$ can
``propagate'' to the right and change the state of the next
neighbor are $U_{x}\left(\begin{array}{c} 1\\[-0.0cm]0
\end{array}\right)
=\left(\begin{array}{c} 0\\[-0.0cm]1 \end{array}\right)U_{x+1}$ and
$D_{x}\left(\begin{array}{c} 0\\[-0.0cm]1 \end{array}\right)=\left(
\begin{array}{c} 1\\[-0.0cm]0 \end{array}\right)D_{x+1}$. So, in order
for the operator $U_1$ to be able to change the configuration of a pair
at $x$, all $x-1$ pairs to the left of this pair must be in the state 
$\left(\begin{array}{c} 1\\[-0.0cm]0\end{array}\right)$. 
Thus, the probability that a pair at $x$ changes its configuration as a
result of the addition of a grain of sand at the left end of the system
is $1/3^{x-1}$. In other words, the system has an exponentially long
configuration memory. If the system is driven by a random sequence of
$U_1$ and $D_1$, on average it will take $\tau_x \sim (3^{x-1})^2 =
9^{x-1}$ grains of sand to change the configuration of a pair at $x$.

\begin{figure}[h]
\narrowtext
\centerline{\epsfxsize=3.3in
\epsffile{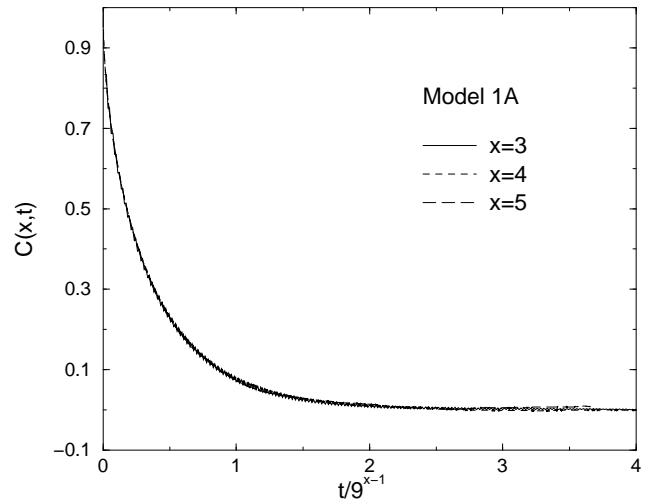}}
\caption{Autocorrelation functions $C(x,t)$ at $x=3$, $x=4$, and $x=5$,
for Model 1A with $L_x=8$.}
\label{auto-di}
\end{figure}

This exponential increase of the characteristic time $\tau_x$ with $x$
is manifested in the local autocorrelation functions
$C(x,t)=[<Z(x,0)Z(x,t)>-<Z(x,0)>^2]/[<Z(x,0)^2>-<Z(x,0)>^2]$, where 
$Z(x,t)=z(x,1)+z(x,2)$ is the number of grains at column $x$ and at time
$t$. In Fig.~\ref{auto-di} we plot $C(x,t)$ {\it vs.} $t/9^{x-1}$ for
several $x$'s. One sees that $C(x,t)=F(t/9^{x-1})$. This form of the
local autocorrelation function implies a scaling form for the local
power spectrum: $S_{loc}(f,x)=(1/f_{char}(x)) S(f/f_{char}(x))$ with
$f_{char}(x)=f_0 \exp (-x\ln 9)$, where $S_{loc}(f,x)$ is the power
spectrum of $Z(x,t)$. Note that (see Eq.~(\ref{u10})) if $U_x$ or
$D_x$ propagates through a column it leaves the number of grains on
that column $Z(x)$ unchanged. It follows that the addition of a grain
at the left end of the system can change at most the number of grains
at one column. If we assume that the local events of changing $Z(x)$
are independent for different $x$'s (or the correlations of which is
not too strong), which is a reasonable approximation when we drive
Model 1A with a random sequence of $U_1$'s and $D_1$'s, then the
global power spectrum of the total number of grains in the system is
the superposition of the local power spectrum.
The exponential fall off of the local characteristic frequencies of
configuration changes would give rise to a global $1/f$ power spectrum,
similar to the case of thermal activations in equilibrium systems
mentioned earlier \cite{dutta}. That is $S_g(f)= \int_{0}^{L}
S_{loc}(f,x) dx = \int_{0}^{L} (1/f_{char}(x)) S(f/f_{char}(x)) dx =
\int_{0}^{L} \exp(\lambda x) S(f \exp (\lambda x)/f_0) dx/f_0 =(1/f)
\int_{f/f_0}^{f \exp(\lambda L)/f_0}$ $dy S(y)/\lambda$.
The lower cutoff of the $1/f$ region is $f_c \sim f_0 \exp (-\lambda
L)$, which for the curve (a) in Fig.~\ref{pw-di} ($\lambda=\ln 9$ and
$L=8$) is of the order $10^{-7}$.

\section{Undirected models}
\begin{figure}[h]
\narrowtext
\centerline{\epsfxsize=3.3in
\epsffile{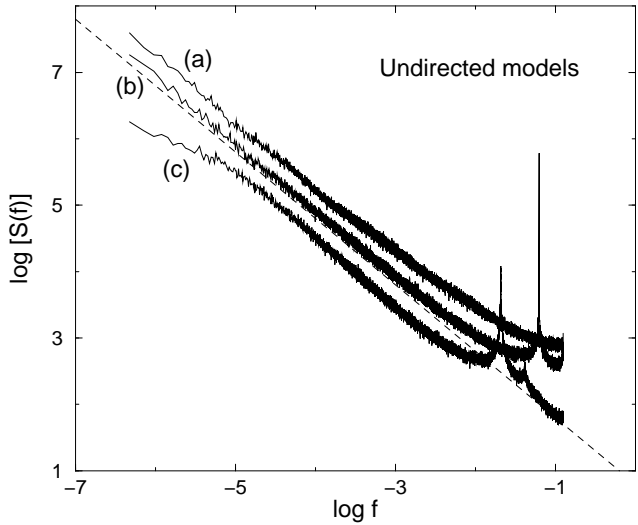}}
\caption{Power spectra for the center-driven BTW models: (a) Model 2A
with $L_x=16$, (b) Model 2 with $L_x=17$ and $L_y=8$, and (c) Model 2
with $L_x=17$ and $L_y=24$. The dashed line has the slope $-1$.}
\label{pw-btw}
\end{figure}

We now turn our attention to the undirected sandpile model on a stripe
$L_x \times L_y$. In this model an unstable site with $z(x,y)>z_c=3$
redistributes one grain of sand to each of its four neighbors.
In our simulations at each time step we randomly select a site 
on the central column ($x=(L_x+1)/2$, or, if $L_x$ is an even number we
randomly select one of the $2 L_y$  sites on the two central columns)
and add one grain of sand to that site. We choose to have open
boundaries at $x=0$ and $x=L_x+1$. In the first version of the
model the boundary condition along the $y$ direction is periodic. 
We refer to this version as Model 2. In addition to Model 2 we
consider a simpler model which we refer to as Model 2A. Model 2A is
defined on an $L \times 2$ stripe with closed boundary condition in 
the $y$-direction. In Model 2A $z_c=2$ and a site with $z(x,y)>2$ moves
one grain of sand to each of its three neighbors. The advantage of
studying Model 2A is that it has fewer recurrent
configurations which are easier to classify.
In Fig.~\ref{pw-btw}, we show the power spectra of the total amount
of sand in Models 2 and 2A. Similar to the case of the directed Model 1,
in Model 2 the $1/f$ region shrinks for increasing $L_y$. The dynamics
of the undirected models is apparently more complex than that of the
directed ones. However, much of the apparent complexity is due to the
motion of ``troughs'' \cite{java} -- columns in which all $z \leq z_c-1$, 
so that avalanches can not propagate beyond them \cite{trough}. 
Let us first understand the trough dynamics. 

\subsection{Trough dynamics}
Let us focus on Model 2A. Like in Model 1A, one can define operators
$U_x$ and $D_x$. 
The operator algebra now satisfies the following operator relations:
\begin{eqnarray}
U_x^3 &=& D_x U_{x+1}U_{x-1} \label{u_btw}\\
D_x^3 &=& U_x D_{x+1}D_{x-1}
\label{d_btw}
\end{eqnarray}
The open boundaries at two ends imply $U_0=D_0=U_{L+1}=D_{L+1}={\bf I}$,
where, as before, ${\bf I}$ is the identity operator. Let us define 
the operator $O_x=U_x D_x$. In Model 1A we have shown that $O_x={\bf I}$
for every $x$. This is not so in the undirected model. However, in this
model the operators $O_x$ form a simple small subgroup of all operators
in the system. From Eqs.~(\ref{u_btw}) and (\ref{d_btw}) it follows that
$O_x^2=O_{x+1}O_{x-1}$. 
Using this operator identity repeatedly one gets
$O_x^2=O_{x+1}O_{x-1}=O_{x+2}O_{x-2}=\ldots=O_{x+n}O_{x-n}$. Similarly,
one can show that 
$O_x O_{x'}=O_{x-1} O_{x'+1}=O_{x-2} O_{x'+2}= \ldots=O_0 O_{x+x'}=
O_{x+x'}$, for $x+x' \leq L$. 
In general,
\beq
O_x O_{x'} = O_{(x+x') \bmod (L+1)}.
\label{subgroup}
\eeq
In other words, operators $O_0 (={\bf I}),O_1,O_2, \ldots, O_L$
form a cyclic subgroup of $L+1$ elements \cite{footnote1}. 

To understand the physical nature of this subgroup let us take a closer
look at the set of recurrent configurations in Model 2A. In a stable
configuration each $z(x,y)$ can take values $0,1,2$. Each pair thus has
9 stable configurations. However, the number of recurrent configurations
is much smaller than $9^L$. To check which stable configurations are
forbidden in the recurrent set one applies the rule developed in
\cite{dhar}. According to this rule, a subconfiguration at a subset of
sites $F$ is forbidden if for every site $(x,y) \in F$ $z(x,y)$ is
strictly smaller than the number of its neighbors {\it in the subset
}$F$.
It is clear that the pair \pair{0}{0} is forbidden. Let us refer to
pairs
\pair{1}{1}, \pair{0}{1}, \pair{1}{0} as {\it troughs}. Troughs prevent
any avalanche from propagating beyond them. It is easy to see that a
subconfiguration enclosed by two troughs is forbidden. Thus a recurrent
configuration cannot contain more than one trough. Therefore, all SOC
states naturally fall into one of the $L+1$ classes: those with no
troughs,
and those with a trough in the $m$-th column. The operator $O_m$ acting
on
a state $|S_m>$ with a trough at the position $m$ does not produce any
topplings but simply fills up the trough. On the other hand, acting on
this state with $O_k$ ($k \neq m$) creates a (usually large) avalanche
in which two grains of sand fall of the pile. However, this avalanche
produces only minor changes in the configuration of the pile. Indeed,
since $O_k=O_m O_{(m-k) \bmod (L+1)}^{-1}$ the operator $O_k$ acting on
the state $|S_m>$ fills up the trough at $x=m$ and creates a trough at
$x=(m-k) \bmod (L+1)$ \cite{footnote2}.
One may simply view that $O_k$ moves the trough from $x=m$ to $x=m-k 
\bmod (L+1)$. The action of $O_k$ on a state $|S>$ with no troughs
results
in a system-wide avalanche with 4 grains of sand falling off the
sandpile.
The only configurational change, however, is the creation of a new
trough
at $L+1-k$ (recall the operator identity $O_k=O_{L+1-k}^{-1}$). These
rules
mean that the action of the $L$ operators $O_k$ results only in the
motion
(or creation and annihilation) of the trough, and does not destroy the
configuration memory of the system. 

\subsection{The number of recurrent configurations}
Having understood that operators $O_k=U_k D_k$,
do not destroy the long term configuration memory of the system
but only move, create, and annihilate troughs, we proceed with
describing how individual operators $U_k=D_k^{-1} O_k$ change
the configuration. We separate the trivial trough dynamics from others
by
defining the equivalence relation of operators: if $A=BO_k$, we say that
$A$ is equivalent to $B$ and denote it by $A\cong B$. Thus
$U_k\cong D_k^{-1}$. One can rewrite the basic operator identity
(\ref{u_btw}) as $U_k^4=U_{k-1} U_{k+1} O_k$, or
\beq
U_k^4\cong U_{k-1} U_{k+1}. 
\label{u4}
\eeq
We now derive the relation between operators $U_k$ at different
sites. Since $U_0={\bf I}$, Eq.~(\ref{u4}) implies $U_1^4 \cong U_2$. 
Write $U_1^{N(k)} \cong U_{k+1}$. One has $U_1^{4N(k)} \cong 
U_{k+1}^4 \cong U_{k}U_{k+2} \cong U_1^{N(k-1)}U_1^{N(k+1)}$, which
gives the recursion relation $N(k+1)=4 N(k)-N(k-1)$ with initial
conditions
$N(0)=1$, $N(1)=4$. It is easy to show that 
\beq
N(k)=[(3+2 \sqrt{3})(2+\sqrt{3})^k - (2 \sqrt{3}-3)(2-\sqrt{3})^k]/6.
\label{nk}
\eeq
In a system of size $L$ one has $U_1^{N(L)} \cong U_{L+1}={\bf I}$.
This observation enables us to calculate the total number of recurrent
(SOC) states as $N_{SOC}^{(2A)}=(L+1) N(L)$, or
\beqar
N_{SOC}^{(2A)}=&{L+1 \over 6}&[(3+2 \sqrt{3})(2+\sqrt{3})^L 
\nonumber \\
&-&(2 \sqrt{3}-3)(2-\sqrt{3})^L].
\label{n_soc_m2a}
\eeqar
In other words, any recurrent configuration can be obtained from a given
one by the action of some power of $U_1$ (there are N(L) choices of this
power before configurations start repeating themselves) and, if
necessary, creation, annihilation, or change of the position of the 
trough achieved by the the action of $L$ operators
$O_k$. \cite{footnote3}.
Asymptotically, only $2+\sqrt{3} \simeq 3.732$ pair configurations 
per site are allowed in a recurrent state, compared to $9$
stable pair configurations. 
The above operator relations can be easily modified for Model 2 on an
$L \times 2$ lattice. For this model the subgroup \req{subgroup} 
of operators $O_{k}$ remains unchanged, while \req{u4} becomes 
$U_k^6\cong U_{k-1} U_{k+1}$. The number of recurrent configurations 
in this model is given by  
$N_{SOC}^{(2)}=(L+1)[(4+3 \sqrt{2})(3+2\sqrt{2})^L 
-(3 \sqrt{2}-4)(3-2 \sqrt{2})^L]/8$.

\subsection{Configuration memory}
\begin{figure}[h]
\narrowtext
\centerline{\epsfxsize=3.3in
\epsffile{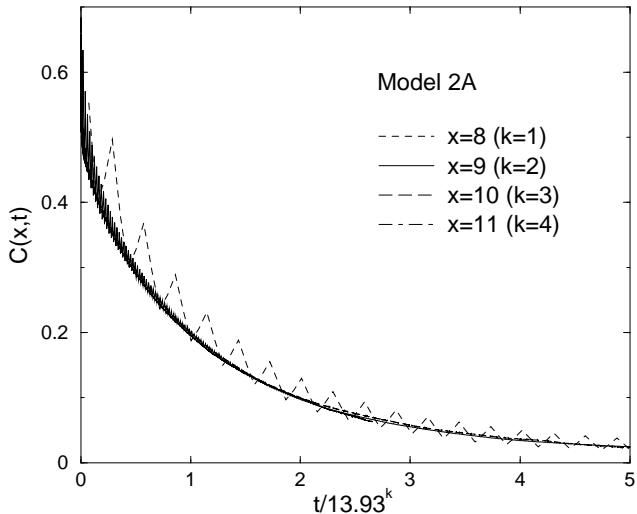}}
\caption{Autocorrelation functions $C(x,t)$ for Model 2A with $L_x=12$,
at $x=8, 9, 10, 11$, which are $k=1, 2, 3, 4$ distance away from the
driving pairs $x=6$ and $x=7$.}
\label{auto-btw}
\end{figure}

Now we are in the position to address the question of long memory in 
Model 2A. (For Model 2 on an $L \times 2$ stripe these arguments 
can be repeated step by step with some minor changes.)
Let us restrict ourselves to the operator $U_k$ acting on a state that
has no trough. Indeed, since we are interested in general properties of
the equivalency class produced by the action of $O_k$'s, one can alway
select from this class a representative state that has no trough.
The action of the operator $U_k$ on the pair at column $k$ is given by
\beqar
U_k\left(\begin{array}{c} 0\\[-0.0cm]2 \end{array}\right)&=&
\left(\begin{array}{c} 1\\[-0.0cm]2 \end{array}\right), \\
U_k\left(\begin{array}{c} 1\\[-0.0cm]2 \end{array}\right)&=&
\left(\begin{array}{c} 2\\[-0.0cm]2 \end{array}\right), \\
U_k\left(\begin{array}{c} 2\\[-0.0cm]0 \end{array}\right)&=&
\left(\begin{array}{c} 0\\[-0.0cm]1 \end{array}\right)U_{k+1}U_{k-1},
\label{u20} \\
U_k\left(\begin{array}{c} 2\\[-0.0cm]1 \end{array}\right)&=&
\left(\begin{array}{c} 2\\[-0.0cm]0 \end{array}\right)O_k, \label{u21} \\ 
U_k\left(\begin{array}{c} 2\\[-0.0cm]2 \end{array}\right)&=&
\left(\begin{array}{c} 2\\[-0.0cm]1 \end{array}\right)O_k. \label{u22}
\eeqar
It seems that Eq.~(\ref{u20}) could propagate $U_k$ through
a string of \pair{2}{0}'s, changing the configurations away from the
driving point, similar to the case of directed models. This is not so,
because one can not have consecutive columns of \pair{2}{0} in a
recurrent state. In fact it is Eqs.~(\ref{u21}) and (\ref{u22}) which
can cause configuration changes away from the driving point. Naively,
according to Eqs.~(\ref{u21}) and (\ref{u22}), the action of $U_k$ on
\pair{2}{1} or \pair{2}{2} causes only local changes apart from some
trough dynamics. However, this is true only if the local changes
(\pair{2}{1}$\to$\pair{2}{0} or \pair{2}{2}$\to$\pair{2}{1}) do not
result in any FSC. If an FSC does
appear as a result of this change, the change in the original
configuration will not be restricted to one pair, but instead will
propagate throughout the FSC. It is an easy task to classify all FSC's
in Model 2A. We have already shown that the pair \pair{0}{0} and two
troughs together with the region between them are FSC's. The third
important FSC is a string
$\left(\begin{array}{ccccccc} 
0 & 1 & 1 & \cdots & 1 & 1 & 0 \end{array} \right)$
(or any other string of $1$'s connecting two $0$'s, 
which we disregard for a
moment, since it must contain at least one trough \pair{1}{1}, and 
we currently restrict ourselves to the states without troughs.)
It is easy to check that in such a string $z(x,y)$ everywhere 
is smaller than the number of its neighbors in the FSC.
Such configuration can be created by
the action of $U_k$ on \pair{2}{2}. Usually one has 
$U_k$\pair{2}{2}$\cong
D_k^{-1}$\pair{2}{2}=\pair{2}{1}. However, if the \pair{2}{2} 
happens to be in the string of \pair{2}{1}'s bounded between 
two \pair{2}{0}'s, simply changing \pair{2}{2} to \pair{2}{1} 
would create a FSC. Direct application of the rules of avalanche
dynamics shows that in this case all pairs associated with this FSC
and the pairs next to it will be updated.
The other way of creating the above mentioned FSC is for $U_k$ to act on
a string of pairs of \pair{2}{1} ended with \pair{2}{0}, 
$\left(\begin{array}{cccccc} 2 & 2 & \cdots & 2 & 2 & 2 \\
1 & 1 & \cdots & 1 & 1 & 0 \end{array} \right)$. 
Again, in this case all pairs associated with this FSC and the pairs
next to it will be changed. 
Both of the scenarios require that the starting configuration contains a
string of \pair{2}{1}'s with \pair{2}{0} at least one of the ends.
Such a string of length $x$ is just one among 
$N_{SOC}^{(2A)}(x) \sim (2+\sqrt{3})^x \simeq 3.732^{x}$ 
recurrent states of a string of $x$ columns. That is why the irrevesible
changes of pairs at distance $x$ from the driving pair are exponentially
unlikely. In Model 2A, driven by random addition of sand at sites on the
central pair(s) the characteristic frequency at distance $k$ from the
driving point is given by $f_{char}(k) \sim 1/((2+\sqrt{3})^{k})^2 
= 13.93^{-k}$. This analytical result is in good agreement with
numerical simulations of the model shown in Fig.~\ref{auto-btw}. 

\section{Conclusion}
In spite of the apparent differences between directed and undirected
models, the mechanism for a long term memory and the $1/f$ spectrum in
these two systems is the same. We summarize our analysis as follows:
\begin{itemize}
\item[1)]
Operators $O_x=U_x D_x$ do not produce irreversible changes in 
the configuration. In the directed models these operators do not change
the configuration at all, while in the undirected models their ability
to change the configuration is restricted to creation, annihilation and
motion of the trough. 
\item[2)] 
In order to produce irreversible changes in a configuration
at a distance $k$ from the place of sand addition, all $k$ pairs
in between have to be in a unique peculiar configuration. 
Since this particular subconfiguration is just one among 
$N_{SOC}(k) \sim A^k$ possible recurrent subconfigurations,
the characteristic frequency of irreversible updates 
falls off with distance exponentially. 
\item[3)] 
Such exponential dependence of the characteristic frequency of
updates leads to the $1/f$ spectrum of the total amount of sand in the
sandpile.
\end{itemize}
It is straightforward to generate the case to higher dimensions in which
sand flows in one (say the $x$) direction with closed or periodic
boundaries in other directions. One would still observe the $1/f$
spectrum. 
Recently, De Los Rios and Zhang \cite{paolo_zhang} observed an $1/f$
spectrum in a non-conserved sandpile-like model in which certain
fraction of sand is lost in each toppling process. Due to the absence
of conservation avalanches themselves are exponentially unlikely to
reach a distant site, giving rise naturally to an exponential
distribution of time scales. In contrast, in our model avalanches
constantly pass through the system but they produce only small changes
of the configuration. 
An $1/f$ spectrum was also observed previously for a {\it continuously}
boundary-driven BTW model \cite{jensen2}. Its origin was attributed to
a (linear) diffusion of $z(x,y)$ with a noisy boundary condition
\cite{jensen2,grin}, which gives a power-law lower cutoff $f_c \sim
1/L_x^2$ for the $1/f$ spectrum -- a mechanism very different than ours.

\begin{center}
{\bf ACKNOWLEDGEMENTS}
\end{center}

SM and YCZ would like to thank NEC Research Institute for its warm
hospitality where this work started.


\begin{references}
\bibitem{press}
For a review, see W. H. Press, Comments on Astrophys. {\bf 7}, 103
(1978).

\bibitem{dutta}
P. Dutta and P. M. Horn, Rev. Mod. Phys. {\bf 53}, 497 (1981).

\bibitem{btw}
P. Bak, C. Tang, and K. Wiesenfeld, Phys. Rev. Lett. {\bf 59}, 381
(1987);
Phys. Rev. A {\bf 38}, 364 (1988).

\bibitem{jensen}
H.J. Jensen, K. Christensen, and H.C. Fogedby, Phys. Rev. B {\bf 40},
7425 (1989).

\bibitem{java}
The readers themselves can observe this effect and many other
interesting things by playing with the hands-on Java applets for
these models at
http://cmth.phy.bnl.gov/\~{}maslov/one-over-f.htm.

\bibitem{dhar}
D. Dhar, Phys. Rev. Lett. {\bf 64}, 1613 (1990). 

\bibitem{dhar_review} 
For a recent review, see D. Dhar, cond-mat/9808047.

\bibitem{dhar_ramaswamy}
D. Dhar and R. Ramaswamy, Phys. Rev. Lett. {\bf 63}, 1659 (1989).

\bibitem{trough}
J.M. Carlson, J.T. Chayes, E.R. Grannan, and G.H. Swindle, Phys. Rev. A
{\bf 42}, 2467 (1990); L.P. Kadanoff, A.B. Chhabra, A.J. Kolan, M.J.
Feigenbaum, and I. Procaccia, Phys. Rev. A {\bf 45}, 6095 (1992).

\bibitem{footnote1}
This result is not restricted to Model 2A. In fact, in any BTW model
with
periodic boundary condition along the $y$ direction one can define
$O_x=\prod_{y=1}^{L_y} a(x,y)$, where $a(x,y)$ is the operator of adding
a grain of sand at $(x,y)$. It is easy to see that the $L_x$ operators
defined this way and the identity operator $O_0={\bf I}$ form the cyclic
subgroup.

\bibitem{footnote2}
There is one minor detail one has to take into account
in order for this statement to be exactly true: the action of
$O_k^{-1}$ is a bit more tricky than simply decreasing the height
of both sites in the $k$-th pair by one. For example, what one
should do if the pair is \pair{2}{0} ?  It is easy to see that
$O_k$ \pair{0}{2}=\pair{2}{0}$D_{k+1}D_{k-1}$. Therefore,
$O_{k}^{-1}$\pair{2}{0}$=$\pair{0}{2}$D_{k+1}^{-1}D_{k-1}^{-1}$.
This means that our definition of a trough should be extended to
include configurations like $\left( \begin{array}{ccc}1 & 2 & 1 \\
2 &0 & 2 \end{array} \right)$. In some exotic cases
the operator $O_k^{-1}$ can cause even more changes in the
configuration. The system however does not forget its original
configuration and operators $O_m$ acting on other sites restore
the original pattern of $z(x,y)$ up to the creation of ``true'' troughs
\pair{1}{1}, \pair{1}{0}, and \pair{0}{1}.

\bibitem{footnote3}
One can also calculate $N_{SOC}^{(2A)}$ using Dhar's formula $N_{SOC}
=\det \Delta_{ij}$ \cite{dhar}. A straightforward calculation gives
$N_{SOC}^{(2A)}=\exp [\sum_{k = 1}^{L}( \log [2 - 2\,\cos (\pi \,
k/(L + 1))] + \log [4 - 2\,\cos (\pi \, k/(L + 1))])]$.
It is easy to verify that this formula and Eq.~(\ref{n_soc_m2a}) 
give the same results. 

\bibitem{paolo_zhang}
P. De Los Rios and Y.-C. Zhang, Phys. Rev. Lett., in press, 
(cond-mat/9808343).

\bibitem{jensen2}
H. Jensen, Physica Scripta {\bf 43}, 593 (1991).

\bibitem{grin}
G. Grinstein, T. Hwa, and H. Jensen, Phys. Rev. A {\bf 45}, R559 (1992).


\end{references}
\end{document}